\documentclass[preprint2]{aastex}
\usepackage[varg]{txfonts}
\usepackage{sistyle}
\usepackage{graphicx}
\bibpunct{(}{)}{;}{a}{}{,}

\title{Investigation of systematic effects in \textit{Kepler} data:\\
Seasonal variations in the light curve of HAT-P-7b}

\author{V. Van Eylen, M. Lindholm Nielsen, B. Hinrup, B. Tingley and H. Kjeldsen}
\affil{Stellar Astrophysics Centre, Department of Physics and Astronomy, Aarhus University, Ny Munkegade 120, DK-8000 Aarhus C, Denmark.}
\email{vincent@phys.au.dk}

\shorttitle{Investigation of artefacts in \textit{Kepler} data}
\shortauthors{Van Eylen et al.}

\received{receipt date}
\revised{revision date}
\accepted{acceptance date}

\begin{abstract}
{With years of \textit{Kepler} data currently available, it can now be attempted to measure variations in planetary transit depths over time. To do so, it is of primary importance to understand which systematic effects may affect the measurement of transits. We aim to measure the stability of \textit{Kepler} measurements over years of observations. We present a study of the depth of about 500 transit events of the Hot Jupiter HAT-P-7b, using 14 quarters (Q0-Q13) of data from the \textit{Kepler} Satellite. We find a systematic variation in the depth of the primary transit, related to quarters of data and recurring yearly. These seasonal variations are about 1\%. Within seasons, we find no evidence for trends. We speculate that the cause of the seasonal variations could be unknown field crowding or instrumental artifacts. Our results show that care must be taken when combining transits throughout different quarters of \textit{Kepler} data. Measuring the relative planetary radius of HAT-P-7b 
without taking these systematic effects into account leads to unrealistically low error estimates. This effect could be present in all \textit{Kepler} targets. If so, relative radius measurements of all Hot Jupiters to a precision much better than 1\% are unrealistic.}
\end{abstract}

\keywords{planets and satellites: fundamental parameters --- stars: fundamental parameters --- stars: individual (HAT-P-7) --- stars: oscillations}

\begin{document}

\maketitle

\section{Introduction}

The \textit{Kepler Satellite} was launched on March 6, 2009 and operates from an Earth-trailing orbit around the Sun. The satellite continually measures the brightness of about 150,000 stars in a long-cadence mode (sampled every 29.4 minutes), and a subset of stars are sampled in a short-cadence of 58.8 seconds \citep{borucki2008}. The primary scientific goal of the \textit{Kepler} Mission is to discover transiting exoplanets, identifying 2740 exoplanet candidates to date \citep{borucki2011,batalha2012}.

The photometry from the mission can be used to characterize many parameters of a star-planet system, e.g. star-to-planet radius ratio, planetary inclination angle, semi-major axis and stellar limb darkening coefficients. For close-in Hot Jupiter planets, the phase curve and planetary occultation may also contain information about the planetary emission and reflection \citep[e.g.][]{borucki2009,vaneylen2012} and in some cases even about ellipsoidal variations and Doppler beaming \citep{welsh2010,jackson2012}. Moreover, if the data is short-cadence, the photometry can also be used to study the host star using asteroseismology \citep{christensendalsgaard2010,vaneylen2012}.

Previous studies show that the various system parameters become better constrained as the amount of data increases \citep[e.g.][]{vaneylen2012,morris2013}. In this paper, we investigate the limits to this ever-increasing accuracy. Ultimately, the accuracy with which parameters can be determined depends on the systematic noise sources, including the instrumental stability of the satellite. This is an important line of research, because such systematic changes could in principle be mistaken for actual changes in the star-planet system, e.g. changes in the planetary atmosphere. We propose therefore to probe these limits by tracking changes in transit depth. Given that \textit{Kepler} rotates $90^\circ$ every three months, with each new pointing assigned a new quarter (Q) number and each orientation a season (every fourth quarter), we explore the changes with this in mind, as these pointing changes are the most likely source of systematic noise.

The HAT-P-7 system  \citep{pal2008} is an ideal target for the study of transit variations. It is a relatively bright star (apparent magnitude of $m_\text{V} = 10.5$), orbited by a Hot Jupiter, whose short period (2.2 days) and large radius (1.4 R$_\mathrm{J}$) result in many deep transits -- all observed with short-cadence. In this paper, we analyze \textit{Kepler} data (Q0-Q13) for this exoplanet, searching for significant quarter-by-quarter or seasonal variations.
 
In Section \ref{sec:observations}, we describe the data reduction procedure. In Section \ref{sec:results}, we present our results. We discuss these results in Section \ref{sec:discussion}, exploring possible causes and
testing to see if our results are specific or general. Finally, we conclude in Section \ref{sec:conclusions}.

\section{Observations\label{sec:observations}}

We choose to start with the raw \textit{Kepler} data, despite the existence of the Presearch Data Conditioned data, desiring to tailor the analysis to the peculiarities of this system. We detrend the raw light curve following three steps: (i) we identify all transit-like features and temporarily mask them off, (ii) fit a fourth-order polynomial to the remaining data in one-month sections (this choice is somewhat arbitrary, as using third- or fifth-order polynomials produces no significant differences), then (iii) remove the long-term trends by dividing the data (including the transits) by the fitted curve. This results in a normalized, detrended light curve, which we can then clean for short-term instrumental effects such as outliers, jumps or drifts with local median filtering. We use this cleaned light curve in the subsequent analyses -- first to get the global parameters, then using these global parameters to search for variations.

\section{Results\label{sec:results}}

%

\subsection{Global transit modeling \label{sec:global_fitting}}

Assuming a perfectly Keplerian orbit, the orbital period is found by applying a linear fit to the centers of the individual transits. The resulting orbital period is \SI{2.20473540(9)}{days}. This value is in agreement with other recent period determinations \citep[e.g.][]{vaneylen2012}, consistent with and refining values based on more limited datasets \citep[e.g.][]{borucki2009hatp7}. We test for deviations from linearity \citep[transit timing variations, see e.g.][]{mazeh2013}, but find no evidence. We place an upper limit of just 3 seconds, as the strongest possible amplitude of a sinusoidal variation.

\begin{figure}[ht]
\includegraphics[width=\columnwidth]{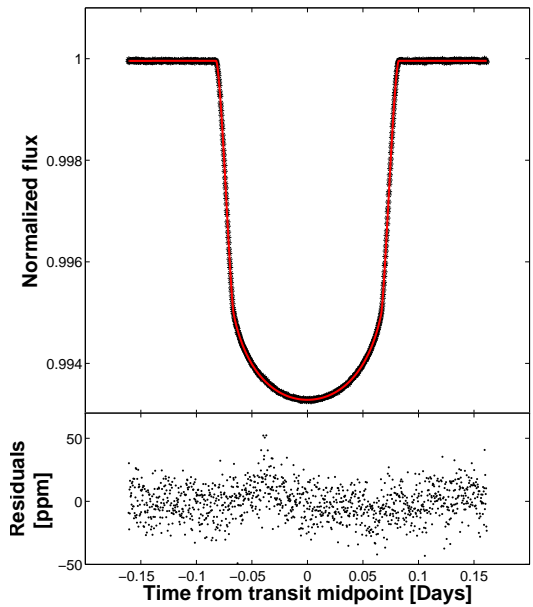}
\caption{\textit{Upper}: Phase folded and binned data near the transit of HAT-P-7b. Individual bins are shown as black dots and the best fit model from TAP is shown in red. \textit{Lower}: Residuals between the binned data and the TAP model.}
\label{fig:bin_fit}
\end{figure}

Having established the absolute periodicity of HAT-P-7, we used the Transit Analysis Package (TAP) \citep{gazak2012} on the phase-folded data, binned to a temporal resolution of \SI{\sim 20}{s}. TAP uses Markov Chain Monte Carlo (MCMC) simulations to fit the transit signature with the analytical methods described in \citet{mandel2002}. When fitting, we used no priors for the limb darkening (LD) coefficients; they were treated as free parameters. The transit light curve and the corresponding fit are shown in Figure~\ref{fig:bin_fit}, with the resulting global planetary parameters listed in Table \ref{tab:primary_best_fit}. These values are in excellent agreement with \cite{esteves2013}. The limb darkening parameters can be compared to predictions from stellar atmospheres, even though they are known to disagree \citep{howarth2011}. Using the stellar parameters by \citet{vaneylen2012}, linear and quadratic values of respectively 0.3201 and 0.4297 are predicted \citep{neilson2013}, for spherically symmetric 
model atmospheres. The residuals from the transit fit (Fig.~\ref{fig:bin_fit}) exhibit an asymmetric trend, also visible in \cite{esteves2013}. This may be due to planet-induced gravitational darkening, as suggested by \cite{morris2013}. 

\begin{table}[ht]
\caption{Summary of all parameters for HAT-P-7b.}
\label{tab:primary_best_fit}
\renewcommand{\arraystretch}{1.2}
\centering
\begin{tabular}{lc}
\hline
Parameter							&	Value\\
\hline
Period P (days)						& 2.20473540 (9)\\
Inclination i ($^\circ$) 						& 83.151 $^{+0.030}_{-0.033}$\\
Rel. semi-major axis a/$R_\star$ 						& 4.1547 $^{+0.0040}_{-0.0042}$\\
Rel. radius $R_\mathrm{p}/R_\star$ 	& 0.077462 $^{+0.000034}_{-0.000034}$\\
  Linear Limb Darkening 						& 0.3440 $^{+0.0064}_{-0.0065}$\\
    Quad. Limb Darkening 						& 0.1843 $^{+0.0098}_{-0.0096}$\\
\hline
\end{tabular}
\end{table}

From these parameters a mean stellar density $\bar{\rho}_\star$ can be derived directly \citep{tingley2011}:

\begin{equation}
\bar{\rho}_\star = \frac{3\pi}{GP^2}\left(\frac{a}{R_\star}\right)^3 - \left(\frac{R_\mathrm{p}}{R_\star}\right)^3 \bar{\rho}_\mathrm{p},
\end{equation}

where $\bar{\rho}_\mathrm{p}$ is the average planetary density, $G$ the gravitational constant, and all other symbols as defined in Table \ref{tab:primary_best_fit}. The second term on the right hand side is often neglected, since the planetary radius is so much smaller than the stellar radius, however we include it in this case. We find $\bar{\rho}_\star = 0.2787\pm0.0008$ g~cm$^{-3}$, with the second term contributing $0.0004$ g~cm$^{-3}$. This is consistent with stellar densities obtained with asteroseismology: $0.2712\pm0.003$ g~cm$^{-3}$ \citep{christensendalsgaard2010} and $0.2781\pm0.0017$ g~cm$^{-3}$ \citep{vaneylen2012}.

\subsection{Transit depth variations}

\begin{figure*}[htbp] 
\centering
\vspace{-1em}
\includegraphics[width = 1\textwidth]{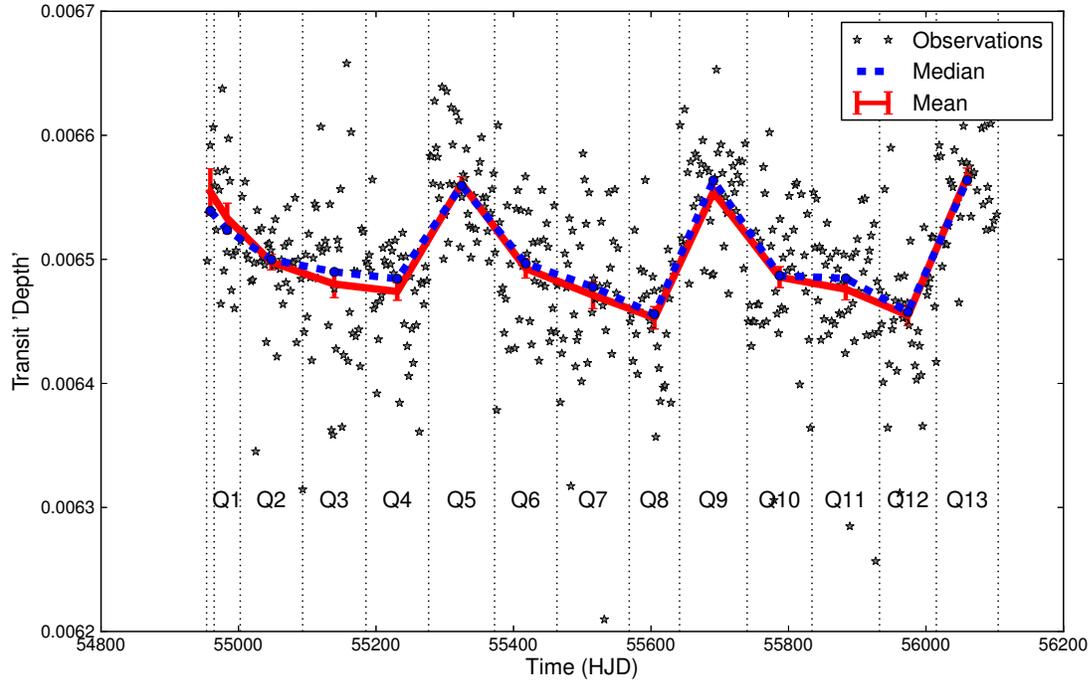}\\
\vspace{-1em}
\includegraphics[width=\columnwidth]{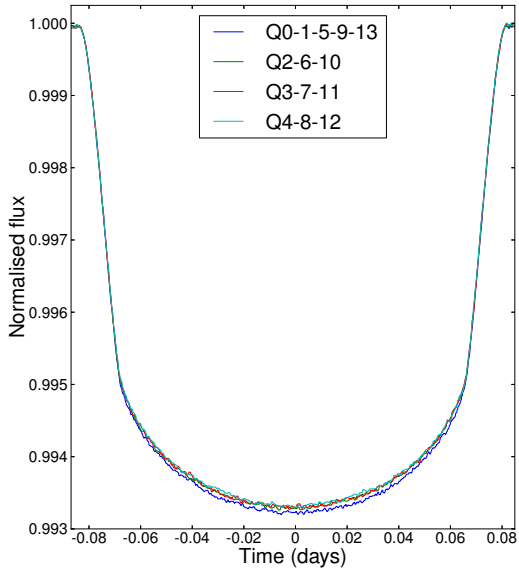}
  \includegraphics[width=\columnwidth]{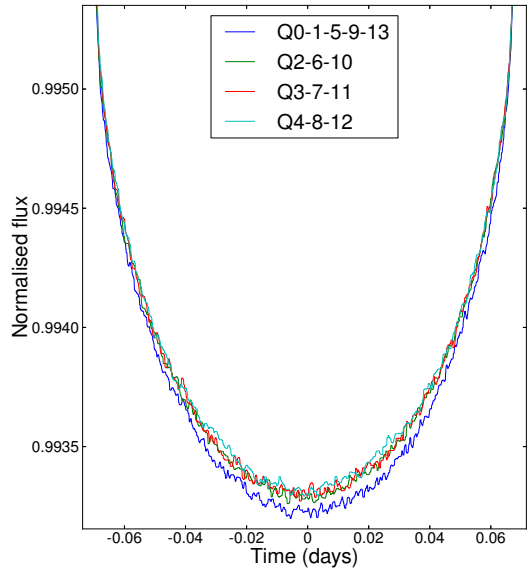}
\caption{\textbf{Top:} Transit depth per quarter, as measured directly from averaging in-transit data points. The measured depth of individual transits (blue dots) is plotted. The green dotted line shows the median per quarter. The red solid line shows the mean and standard deviation per quarter, calculated from non-parametric bootstrapping of the observations. 
\textbf{Bottom:} Transit depth as binned per season. From the left figure, it is clear that the different seasons show the same transit shape. The right figure (zoom) shows the change in depth for different seasons. The observations were binned for clarity.\label{fig:transitdepth}}
\end{figure*}

Having obtained the global transit depth, we now measure the depth of individual transits. The depth as measured here is defined (somewhat arbitrarily) as the average of the in-transit obserations within 65 minutes from the transit center, compared with the average flux level at the transit wings (taking the same amount of data points in and out of transit). We find that the transit depth is dependent on the data quarter, as can be seen in Figure~\ref{fig:transitdepth}. The figure shows the relative depth measurements for individual transits, as well as the average transit depth per quarter of \textit{Kepler} data. To avoid making assumptions about the distribution of the measurements, the error bars are calculated using a bootstrap procedure. A variation in transit depth is clearly visible, with a recurring trend every fourth quarter (every year) -- a seasonal variation. To ensure the crude way of estimating a transit depth is adequate, we also show a quarterly median in Figure~\ref{fig:transitdepth} (top 
panel).

\begin{table*}[th]
\caption{HAT-P-7b measurements per season. The transit `depth' is measured directly as an average of in-transit data points, the relative and per cent depths are calculated by dividing by the average `depth' (6500.7 ppm). $R_\mathrm{p}/R_\star$ is calculated from fitting a physical model to the transits, allowing only this parameter to vary.}
\label{tab:transit_depth}
\renewcommand{\arraystretch}{1.2}
\centering
\begin{tabular}{lllllll}
\hline
Season	& Module & Channel & `Depth' (ppm) & Rel. `depth' & `Depth' diff. (\%)  &	$R_\mathrm{p}/R_\star$						\\
\hline
 1 	& 17 & 58			& 6557.2 $\pm$ 3.5	& 1.00869 $\pm$ 0.00054	 & 0.869$\pm$ 0.054 &	0.077618 $^{+0.000073}_{-0.000073}$ 	\\
 2 	& 19 & 66			& 6491.8 $\pm$ 4.5 	& 0.99863 $\pm$ 0.00068	 & -0.137$\pm$ 0.068 &	0.077355 $^{+0.000035}_{-0.000035}$ 	\\
 3 	& 9 & 26			& 6475.9 $\pm$ 6.0	& 0.99618 $\pm$ 0.00091	 & -0.382$\pm$ 0.091 &	0.077330 $^{+0.000056}_{-0.000059}$ 	\\
 4 	& 7 & 18			& 6461.1 $\pm$ 4.7	& 0.99392 $\pm$ 0.00071	 & -0.608$\pm$ 0.071 &	0.077229 $^{+0.000044}_{-0.000043}$ 	\\
\hline
\end{tabular}
\caption{Measurements of pulsation amplitude of FN Lyr (KIC 6936115) per season. The relative difference and per cent difference are calculated by dividing by the average RMS amplitude (268249 ppm).}
\label{tab:reference_star}
\centering
\begin{tabular}[width = 1\textwidth]{llllll}
\hline
Season & Module & Channel & RMS Amplitude (ppm) & Relative difference & Difference (\%) \\
\hline
1   &     8      &23  &     270723 $\pm$  176    &1.0092 $\pm$ 0.0007 &  0.92 $\pm$ 0.07\\
2    &   12     & 39   &    268014 $\pm$  109   & 0.9991 $\pm$ 0.0004  &-0.09 $\pm$ 0.04\\
3     &  18    &  63    &   266843 $\pm$ 312  &  0.9948 $\pm$ 0.0012 & -0.52 $\pm$ 0.12\\
4      & 14   &   47     &  267414 $\pm$ 1164 &   0.9969 $\pm$ 0.0043 & -0.31 $\pm$ 0.43\\
\hline
\end{tabular}
\end{table*} 
Subsequently, we split the data into four sets of quarters (seasons): a dataset for Q0-1-5-9-13, for Q2-6-10, for Q3-7-11 and finally for Q4-8-12. The resulting average transits, rebinned for clarity, are shown in Figure~\ref{fig:transitdepth}. This figure confirms that the transit depth is dependent on the season of observation. The figure also shows that the \textit{shape} of the transit is seemingly unaffected.

The average depth per quarter is also shown in Table \ref{tab:transit_depth}, as well as the average depth per quarter divided by the average depth for the combined dataset, showing differences of 1.5\% between Season 1 and Season 4, with differences of at least 3$\sigma$ between any two seasons. To investigate this effect further, we fit the individual seasons using the TAP as described in Section \ref{sec:global_fitting}, but only allowed the planetary radius to be free, fixing all other parameters to our earlier derived values. The results are also shown in Table \ref{tab:transit_depth}. For verification purposes, we also performed this analysis on the PDC \textit{Kepler} light curve and found similar results.

\section{Discussion\label{sec:discussion}}

\subsection{The true planetary radius}

While Figure \ref{fig:transitdepth} shows a clear change in depth per \textit{Kepler} quarter, the impact on the planetary radius is less severe (see Table \ref{tab:transit_depth}). The reason for this is two-fold. Firstly, using a full physical description to model the transit increases the size of the error bars than simply measuring the average of all in-transit data points. As a result, the radius difference is statistically less significant than the depth difference. The other reason is the fact that the radius scales with the square root of the transit depth, and is therefore less affected.

Even so, when comparing Season 1 to Season 4, the radius found for HAT-P-7b is different by 6$\sigma$. It is not immediately clear which season is the most accurate. While the true depth may be somewhere in the middle of all seasonal depths, it is also possible that one specific season represents the correct transit depth - or none at all. Indeed, one might speculate that all of the seasons are to some extent affected by this apparent systematic error. The true transit depth can also be important for studies of exoplanet atmospheres, where a different radius is expected in different wavelengths as a consequense of haze or dust \citep[e.g.][]{pont2008}. The reported effect shows care must be taken when comparing \textit{Kepler} values with measurements in other wavelengths. Regardless of the true planetary size, it is clear from Table \ref{tab:primary_best_fit} that the resulting value from a na\"ive fit to the combined \textit{Kepler} photometry ($R_\mathrm{p}/R_\star$ = 0.077462 $\pm$ 0.000034) 
underestimates the true error bars of the relative planetary radius by an order of magnitude.

\subsection{Other targets}

A full comparison of the seasonal variations of the \textit{Kepler} satellite is outside the scope of this paper. Even so, we attempted to confirm this effect in other targets.

The non-Blazhko ab-type RR Lyr star KIC 6936115, also known as FN Lyr \citep{nemec2011} is a very stable pulsating star. Its pulsation amplitude (which should be constant) was therefore used to search for seasonal
dependences in \textit{Kepler} data. The RMS amplitude is calculated from the raw time series. Instead of extracting amplitudes from Fourier Analysis by measuring the amplitude of the different oscillation modes we calculated the RMS amplitude directly from the time series. First, we corrected for the slow sensitivity variation by calculating the moving median of the time series and divide the time series with the moving median series, then we calculated the average RMS amplitude for each quarter (Q0-Q14) and season (shown in Table \ref{tab:reference_star}). The relative amplitude of the pulsations differs from season to season by up to 1\%
with a high level of significance. It is clear from the table that the seasonal variations in this target are real: the offset of Season 1 is separated by 13$\sigma$ from the average. As with HAT-P-7, we have repeated the analysis, replacing the raw \textit{Kepler} data by PDC processed data. The level of our error bars increases slightly (the PDC pipeline is not normally used for studying pulsation amplitudes), but our findings remain intact.

The seasonal differences seem to agree quantitatively with our findings for HAT-P-7b's transit depth, with peak-to-peak variations of 1.5 per cent. It is tempting to interpret this as proof that these seasonal variation occur in all \textit{Kepler} targets. However, it turns out to be complicated to confirm: testing relative stability to a percentage level requires physical events that are stable to this level over a few years. Most stellar pulsations do not fall into this category, and most planetary transits are not as deep or as frequently occurring as HAT-P-7b's. 

One alternative to planetary transits are transits caused by eclipsing binary stars, which have even deeper transits than HAT-P-7b. \citet{slawson2011} mention, for example, that potential quarter-to-quarter systematics may be present in their catalogue of 2165 \textit{Kepler} binary stars. However, the stability of transit depths for binary objects is often not at percentage level, due to various physical phenomena. 

\subsection{Possible causes\label{sec:causes}}

We speculate on four possible causes for seasonal variations: the first two are systematic errors that are caused by crowding in the field and the applied mask, the other two are related to non-linear behavior of the \textit{Kepler} CCD or its color dependence.

(i) \textit{Field crowding} could potentially explain differences in relative measurements. As the \textit{Kepler} CCDs make a roll maneuver at the end of each quarter, the targets fall on different CCDs and a new optimal aperture is applied. Any nearby targets could introduce flux, which could be different every season. The Pre-search Data Conditioning (PDC) attempts to correct for crowding effects by estimating a crowding metric, the fraction of light that comes from the actual target star. This is estimated from the distribution of surrounding objects in the \textit{Kepler} Input Catalogue (KIC). The crowding metric is calculated per quarter and therefore in principle has the capability to correct for the observed effects. We have tested our analysis of both HAT-P-7 and FN Lyr on PDC data and found similar results as when using the raw data. However, this does not rule out the possibility of additional crowding of unknown or undetected sources that the PDC data does not correct for. 

A nearby target, 4-5 \textit{Kepler} magnitudes fainter than HAT-P-7 (\textit{Kepler} magnitude $K_\textrm{v}$ = 10.5), separated by $\sim$~10'' from HAT-P-7 so it is close to the edge of the optimal aperture, could reproduce the observed 1.5 per cent peak-to-peak variation. Such an object should be seen in the UKIRT\footnote{http://keplergo.arc.nasa.gov/ToolsUKIRT.shtml} or UBV \citep{everett2012} catalogues. A target at a distance of 15'' is seen in both catalogues, estimated to be 5.8 (UKIRT) or 7.2 (UBV) $K_\textrm{v}$ fainter. This target appears to be outside the apertures in all seasons, and is presumably too faint to reproduce the observed effect. Two companion stars are observed at about 3'' \citep{narita2010}, but fall inside the aperture during all seasons. The brightest of the two is also seen in UKIRT photometry, and with an estimated 6.5 $K_\textrm{v}$ fainter than HAT-P-7, it is anyhow too faint to produce the seasonal variations. All other known objects are too faint and too distant.

(ii) An alternative explanation related to the target mask is the opposite effect: \textit{a differential loss of photons from the target star}, due to a mask that is too small. A simple loss of target photons would not create a difference in the measured transit depth (this is a relative measurement), however, in principle, it is possible that the photon loss is different in-transit compared to out-of-transit, because the transit dims the brightness of the star, changing the way electrons in saturated pixels overflow into their neighbors. An example of this is the observations of RR Lyr in early \textit{Kepler} photometry, where a mask that was too small to capture the brightest stages of the star, caused an underestimated amplitude \citep{kolenberg2011}. As the dimming due to planetary transits is small, we consider this option rather unlikely. If true, this effect could potentially be mitigated by increasing the size of the mask, assuming there are not too many contaminating sources nearby \citep{
kinemuchi2012}. 

(iii) Another possible explanation for our observations is \textit{non-linearities related to the \textit{Kepler} instrument}. The response to a signal is non-linear to some extent for every CCD and those in \textit{Kepler} are no exception to this rule. The \textit{Kepler} instrument handbook \citep{vancleve2009} states that the observed non-linearity in the \textit{Kepler} focal planes is of order 3\%. A similar value of $\pm$ 3\%, for both saturated and non-saturated targets, is reported based on the first 43 days of \textit{Kepler} observations \citep{caldwell2010}. After a few more months of observations, \citet{caldwell2010b} report that the non-linearity is possibly lower than 3\%, but difficult to measure in-flight. The non-linearity we observe is about 1\%, well within what could be expected based on these predictions, giving credibility to this explanation. One could turn this argument around and state that the \textit{Kepler} instrument is \textit{remarkably stable}. Indeed, the non-
linearity observed is \textit{only 1\%} between seasons, and \textit{essentially full linearity} is measured when observations for the same CCD are compared, even when those observations are years apart.

(iv) A final, albeit somewhat unlikely explanation, is that the effect is caused by a slightly different \textit{color-dependence in the pixel response function (PRF)}. The PRFs for individual \textit{Kepler} channels are available for download. We compared the `average color' of the channels for FN Lyr, as this is known to scale with pulsation amplitude, and found no correlation between color and measured seasonal amplitude, further decreasing the viability of this explanation.

\newpage
\section{Conclusions\label{sec:conclusions}}
We investigated the transit depth of HAT-P-7b for different quarters and find a statistically significant seasonal trend. We have confirmed a similar trend to be present in the amplitude of FN Lyr and speculate this behavior of the \textit{Kepler} data could be caused by instrumental artifacts or by unaccounted-for crowding in the \textit{Kepler} field.

We find the impact on planetary radius measurements to be relatively small, because they scale with the square root of the transit depth: a difference in relative planetary radius of $\pm$ 1\% is reported over different seasons. Nevertheless, these results cast doubt over the accuracy of transit depth measurements of \textit{Kepler} targets, which are often reported at a higher precision than this and may be subject to the same systematic effects reported in this paper. Hot Jupiters are of particular concern, as their relative radii are sometimes reported with error bars that are an order of magnitude smaller than the systematic effect we observe in HAT-P-7 and FN Lyr. Apart from underestimating the error bars, we also point out that, in principle, the correct radius might not be the average of the seasonal measurements -- they might all be affected by a certain degree of systematic error.

If these seasonal variations are indeed persistent throughout \textit{Kepler} measurements, Hot Jupiters such as HAT-P-7b will make them most visible, as they have many deep transit events. When seeking to compare physical variations of transit depths over time (e.g. caused by planetary weather phenomena or variable atmospheres), it is therefore important to be very careful when combining measurements throughout different \textit{Kepler} quarters and seasons. However, planetary transits may be the ideal tool to calibrate and correct for these seasonal systematics.

We finally conclude that within seasons, the \textit{Kepler} instrument is remarkably stable, even for observations that were made years apart.

\acknowledgements
 We thank the anonymous referee for valuable comments, Douglas Caldwell, Mia Lundkvist, Mikkel N. Lund, Rasmus Handberg for help in retrieving relevant reference data, and Simon Albrecht and Jens Jessen-Hansen for valuable discussions. Funding for this Discovery mission is provided by NASA’s Science Mission Directorate. The authors wish to thank the entire \textit{Kepler} team, without whom these results would not be possible. Funding for the Stellar Astrophysics Centre is provided by The Danish National Research Foundation (Grant agreement no.: DNRF106). The research is supported by the ASTERISK project (ASTERoseismic Investigations with SONG and \textit{Kepler}) funded by the European Research Council (Grant agreement no.:~267864).


\begin{thebibliography}{30}
\expandafter\ifx\csname natexlab\endcsname\relax\def\natexlab#1{#1}\fi

\bibitem[{{Batalha} \& {Kepler Team}(2012)}]{batalha2012}
{Batalha}, N.~M. \& {Kepler Team}. 2012, in American Astronomical Society
  Meeting Abstracts, Vol. 220, American Astronomical Society Meeting Abstracts
  220, 306.01

\bibitem[{{Borucki} {et~al.}(2008){Borucki}, {Koch}, {Basri}, {Batalha},
  {Brown}, {Caldwell}, {Christensen-Dalsgaard}, {Cochran}, {Dunham}, {Gautier},
  {Geary}, {Gilliland}, {Jenkins}, {Kondo}, {Latham}, {Lissauer}, \&
  {Monet}}]{borucki2008}
{Borucki}, W., {Koch}, D., {Basri}, G., {et~al.} 2008, in IAU Symposium, Vol.
  249, IAU Symposium, ed. Y.-S. {Sun}, S.~{Ferraz-Mello}, \& J.-L. {Zhou},
  17--24

\bibitem[{{Borucki} {et~al.}(2009{\natexlab{a}}){Borucki}, {Koch}, {Batalha},
  {Caldwell}, {Christensen-Dalsgaard}, {Cochran}, {Dunham}, {Gautier}, {Geary},
  {Gilliland}, {Jenkins}, {Kjeldsen}, {Lissauer}, \& {Rowe}}]{borucki2009}
{Borucki}, W., {Koch}, D., {Batalha}, N., {et~al.} 2009{\natexlab{a}}, in IAU
  Symposium, Vol. 253, IAU Symposium, ed. F.~{Pont}, D.~{Sasselov}, \& M.~J.
  {Holman}, 289--299

\bibitem[{{Borucki} {et~al.}(2009{\natexlab{b}}){Borucki}, {Koch}, {Jenkins},
  {Sasselov}, {Gilliland}, {Batalha}, {Latham}, {Caldwell}, {Basri}, {Brown},
  {Christensen-Dalsgaard}, {Cochran}, {DeVore}, {Dunham}, {Dupree}, {Gautier},
  {Geary}, {Gould}, {Howell}, {Kjeldsen}, {Lissauer}, {Marcy}, {Meibom},
  {Morrison}, \& {Tarter}}]{borucki2009hatp7}
{Borucki}, W.~J., {Koch}, D., {Jenkins}, J., {et~al.} 2009{\natexlab{b}},
  Science, 325, 709

\bibitem[{{Borucki} {et~al.}(2011){Borucki}, {Koch}, {Basri}, {Batalha},
  {Brown}, {Bryson}, {Caldwell}, {Christensen-Dalsgaard}, {Cochran}, {DeVore},
  {Dunham}, {Gautier}, {Geary}, {Gilliland}, {Gould}, {Howell}, {Jenkins},
  {Latham}, {Lissauer}, {Marcy}, {Rowe}, {Sasselov}, {Boss}, {Charbonneau},
  {Ciardi}, {Doyle}, {Dupree}, {Ford}, {Fortney}, {Holman}, {Seager},
  {Steffen}, {Tarter}, {Welsh}, {Allen}, {Buchhave}, {Christiansen}, {Clarke},
  {Das}, {D{\'e}sert}, {Endl}, {Fabrycky}, {Fressin}, {Haas}, {Horch},
  {Howard}, {Isaacson}, {Kjeldsen}, {Kolodziejczak}, {Kulesa}, {Li}, {Lucas},
  {Machalek}, {McCarthy}, {MacQueen}, {Meibom}, {Miquel}, {Prsa}, {Quinn},
  {Quintana}, {Ragozzine}, {Sherry}, {Shporer}, {Tenenbaum}, {Torres},
  {Twicken}, {Van Cleve}, {Walkowicz}, {Witteborn}, \& {Still}}]{borucki2011}
{Borucki}, W.~J., {Koch}, D.~G., {Basri}, G., {et~al.} 2011, \apj, 736, 19

\bibitem[{{Caldwell} {et~al.}(2010{\natexlab{a}}){Caldwell}, {Kolodziejczak},
  {Van Cleve}, {Jenkins}, {Gazis}, {Argabright}, {Bachtell}, {Dunham}, {Geary},
  {Gilliland}, {Chandrasekaran}, {Li}, {Tenenbaum}, {Wu}, {Borucki}, {Bryson},
  {Dotson}, {Haas}, \& {Koch}}]{caldwell2010}
{Caldwell}, D.~A., {Kolodziejczak}, J.~J., {Van Cleve}, J.~E., {et~al.}
  2010{\natexlab{a}}, \apjl, 713, L92

\bibitem[{{Caldwell} {et~al.}(2010{\natexlab{b}}){Caldwell}, {van Cleve},
  {Jenkins}, {Argabright}, {Kolodziejczak}, {Dunham}, {Geary}, {Tenenbaum},
  {Chandrasekaran}, {Li}, {Wu}, \& {von Wilpert}}]{caldwell2010b}
{Caldwell}, D.~A., {van Cleve}, J.~E., {Jenkins}, J.~M., {et~al.}
  2010{\natexlab{b}}, in Society of Photo-Optical Instrumentation Engineers
  (SPIE) Conference Series, Vol. 7731, Society of Photo-Optical Instrumentation
  Engineers (SPIE) Conference Series

\bibitem[{{Christensen-Dalsgaard} {et~al.}(2010){Christensen-Dalsgaard},
  {Kjeldsen}, {Brown}, {Gilliland}, {Arentoft}, {Frandsen}, {Quirion},
  {Borucki}, {Koch}, \& {Jenkins}}]{christensendalsgaard2010}
{Christensen-Dalsgaard}, J., {Kjeldsen}, H., {Brown}, T.~M., {et~al.} 2010,
  \apjl, 713, L164

\bibitem[{{Esteves} {et~al.}(2013){Esteves}, {De Mooij}, \&
  {Jayawardhana}}]{esteves2013}
{Esteves}, L.~J., {De Mooij}, E.~J.~W., \& {Jayawardhana}, R. 2013,
  ArXiv:1305.3271

\bibitem[{{Everett} {et~al.}(2012){Everett}, {Howell}, \& {Kinemuchi}}]{everett2012}
{Everett}, M.~E., {Howell}, S.~B., \& {Kinemuchi}, K. 2012, \pasp, 124, 316-322



\bibitem[{{Gazak} {et~al.}(2012){Gazak}, {Johnson}, {Tonry}, {Dragomir},
  {Eastman}, {Mann}, \& {Agol}}]{gazak2012}
{Gazak}, J.~Z., {Johnson}, J.~A., {Tonry}, J., {et~al.} 2012, Advances in
  Astronomy, 2012

  \bibitem[{{Howarth}(2011){Howarth}}]{howarth2011}
{Howarth}, I.~D. 2011, \mnras, 418, 1165

\bibitem[{{Jackson} {et~al.}(2012){Jackson}, {Lewis}, {Barnes}, {Drake Deming},
  {Showman}, \& {Fortney}}]{jackson2012}
{Jackson}, B.~K., {Lewis}, N.~K., {Barnes}, J.~W., {et~al.} 2012, \apj, 751,
  112

\bibitem[{{Kinemuchi} {et~al.}(2012){Kinemuchi}, {Barclay}, {Fanelli},
  {Pepper}, {Still}, \& {Howell}}]{kinemuchi2012}
{Kinemuchi}, K., {Barclay}, T., {Fanelli}, M., {et~al.} 2012, \pasp, 124, 963

\bibitem[{{Kolenberg} {et~al.}(2011){Kolenberg}, {Bryson}, {Szab{\'o}},
  {Kurtz}, {Smolec}, {Nemec}, {Guggenberger}, {Moskalik}, {Benk{\H o}},
  {Chadid}, {Jeon}, {Kiss}, {Kopacki}, {Nuspl}, {Still},
  {Christensen-Dalsgaard}, {Kjeldsen}, {Borucki}, {Caldwell}, {Jenkins}, \&
  {Koch}}]{kolenberg2011}
{Kolenberg}, K., {Bryson}, S., {Szab{\'o}}, R., {et~al.} 2011, \mnras, 411, 878

\bibitem[{{Mandel} \& {Agol}(2002)}]{mandel2002}
{Mandel}, K. \& {Agol}, E. 2002, \apjl, 580, L171

\bibitem[{{Mazeh} {et~al.}(2013){Mazeh}, {Nachmani}, {Holczer}, {Fabrycky},
  {Ford}, {Sanchis-Ojeda}, {Sokol}, {Rowe}, {Agol}, {Carter}, {Lissauer},
  {Quintana}, {Ragozzine}, {Steffen}, \& {Welsh}}]{mazeh2013}
{Mazeh}, T., {Nachmani}, G., {Holczer}, T., {et~al.} 2013, ArXiv:1301.5499

\bibitem[{{Morris} {et~al.}(2013){Morris}, {Mandell}, \& {Deming}}]{morris2013}
{Morris}, B.~M., {Mandell}, A.~M., \& {Deming}, D. 2013, \apjl, 764, L22

\bibitem[{{Narita} {et~al.}(2010){Narita}, {Kudo}, {Bergfors}, {Nagasawa},
  {Thalmann}, {Sato}, 
  {Suzuki}, {Kandori}, 	{Janson}, {Goto}, {Brandner}, {Ida}, 
	{Abe}, {Carson}, {Egner}, {Feldt},  
	{Golota}, {Guyon}, {Hashimoto}, {Hayano}, {Hayashi}, {Hayashi}, {Henning}, {Hodapp}, 
	{Ishii}, {Knapp}, {Kusakabe}, {Kuzuhara}, 
	{Matsuo}, {McElwain}, {Miyama}, {Morino}, {Moro-Martin}, {Nishimura}, {Pyo}, {Serabyn},  
	{Suenaga}, {Suto}, {Takahashi}, {Takami}, 
	{Takato}, {Terada}, {Tomono}, {Turner}, 
	{Watanabe}, {Yamada}, {Takami}, {Usuda} \& {Tamura}}]{narita2010}
{Narita}, N., {Kudo}, T., {Bergfors}, C., {et~al.} 2010, \pasj, 62, 779

\bibitem[{{Neilson} \& {Lester}(2013)}]{neilson2013}
{Neilson}, H.~R. \& {Lester}, J.~B. 2013, ArXiv:1306.6640

\bibitem[{{Nemec} {et~al.}(2011){Nemec}, {Smolec}, {Benk{\H o}}, {Moskalik},
  {Kolenberg}, {Szab{\'o}}, {Kurtz}, {Bryson}, {Guggenberger}, {Chadid},
  {Jeon}, {Kunder}, {Layden}, {Kinemuchi}, {Kiss}, {Poretti},
  {Christensen-Dalsgaard}, {Kjeldsen}, {Caldwell}, {Ripepi}, {Derekas},
  {Nuspl}, {Mullally}, {Thompson}, \& {Borucki}}]{nemec2011}
{Nemec}, J.~M., {Smolec}, R., {Benk{\H o}}, J.~M., {et~al.} 2011, \mnras, 417,
  1022

\bibitem[{{P{\'a}l} {et~al.}(2008){P{\'a}l}, {Bakos}, {Torres}, {Noyes},
  {Latham}, {Kov{\'a}cs}, {Marcy}, {Fischer}, {Butler}, {Sasselov}, {Sip{\H
  o}cz}, {Esquerdo}, {Kov{\'a}cs}, {Stefanik}, {L{\'a}z{\'a}r}, {Papp}, \&
  {S{\'a}ri}}]{pal2008}
{P{\'a}l}, A., {Bakos}, G.~{\'A}., {Torres}, G., {et~al.} 2008, \apj, 680, 1450

\bibitem[{{Pont} {et~al.}(2008){Pont}, {Knutson}, {Gilliland}, {Moutou}, \& {Charbonneau}}]{pont2008}
{Pont}, F., {Knutson}, H., {Gilliland}, R.~L., {et~al.} 2008, \mnras, 385, 109

\bibitem[{{Slawson} {et~al.}(2011){Slawson}, {Pr{\v s}a}, {Welsh}, {Orosz},
  {Rucker}, {Batalha}, {Doyle}, {Engle}, {Conroy}, {Coughlin}, {Gregg},
  {Fetherolf}, {Short}, {Windmiller}, {Fabrycky}, {Howell}, {Jenkins}, {Uddin},
  {Mullally}, {Seader}, {Thompson}, {Sanderfer}, {Borucki}, \&
  {Koch}}]{slawson2011}
{Slawson}, R.~W., {Pr{\v s}a}, A., {Welsh}, W.~F., {et~al.} 2011, \aj, 142, 160


\bibitem[{{Tingley} {et~al.}(2011){Tingley}, {Bonomo}, \& {Deeg}}]{tingley2011}
{Tingley}, B., {Bonomo}, A.~S., \& {Deeg}, H.~J. 2011, \apj, 726, 112

\bibitem[{{Van Cleve} \& {Caldwell}(2009)}]{vancleve2009}
{Van Cleve}, J.~E. \& {Caldwell}, D.~A. 2009

\bibitem[{{Van Eylen} {et~al.}(2012){Van Eylen}, {Kjeldsen},
  {Christensen-Dalsgaard}, \& {Aerts}}]{vaneylen2012}
{Van Eylen}, V., {Kjeldsen}, H., {Christensen-Dalsgaard}, J., \& {Aerts}, C.
  2012, Astronomische Nachrichten, 333, 1088

\bibitem[{{Welsh} {et~al.}(2010){Welsh}, {Orosz}, {Seager}, {Fortney},
  {Jenkins}, {Rowe}, {Koch}, \& {Borucki}}]{welsh2010}
{Welsh}, W.~F., {Orosz}, J.~A., {Seager}, S., {et~al.} 2010, \apjl, 713, L145

\end{thebibliography}
\end{document}